\documentclass[twocolumn,letter]{jpsj2}
\usepackage[dvips]{graphicx}

\def\runtitle{Pressure-Induced Successive Magnetic Phase Transitions in TlCuCl$_3$}
\def\runauthor{Akira \textsc{Oosawa} {\it et al.}}

\title{Pressure-Induced Successive Magnetic Phase Transitions in the Spin Gap System TlCuCl$_3$}

\author{Akira \textsc{Oosawa}\thanks{E-mail address: a-oosawa@neutrons.tokai.jaeri.go.jp.}, Kazuhisa \textsc{Kakurai}, Toyotaka \textsc{Osakabe}, Mitsutaka \textsc{Nakamura}, Masayasu \textsc{Takeda} and Hidekazu \textsc{Tanaka}$^1$}

\inst{Advanced Science Research Center, Japan Atomic Energy Research Institute, Tokai, Ibaraki 319-1195, Japan\\
$^1$Research Center for Low Temperature Physics, Tokyo Institute of Technology, Oh-okayama, Meguro-ku, Tokyo 152-8551, Japan}

\recdate{\today}

\abst{Under the hydrostatic pressure $P=1.48$ GPa, polarized neutron elastic scattering experiments have been carried out on the coupled spin dimer system TlCuCl$_3$ with the gapped ground state at ambient pressure. Pressure-induced magnetic ordering occurs at the transition temperature $T_{\rm N}=16.9$ K. As in the field-induced and impurity-induced magnetic ordered phases, the ordered moments lie in the $a-c$ plane just below $T_{\rm N}$. An additional spin reorientation (SR) phase transition was observed at $T_{\rm SR}=10$ K, where the ordered moments start to incline toward the $b$-axis. The temperature variations of the direction and the magnitude of the ordered moments were also investigated.}

\kword{TlCuCl$_3$, spin gap, pressure-induced magnetic ordering, successive magnetic phase transitions, spin reorientation, polarized neutron elastic scattering}

\begin{document}
\sloppy
\maketitle

The spin gap in interacting spin systems is a macroscopic quantum phenomenon and has been attracting considerable attention both theoretically and experimentally. Recently, unconventional magnetic orderings induced by impurity ions, applied magnetic field and pressure in the spin gap systems have been energetically investigated. The impurity-induced magnetic ordering due to the breaking of the local spin singlet pair has been well studied through the well-known inorganic spin-Peierls system CuGeO$_3$. The coexistence of the lattice dimerization associated with the spin gap and antiferromagnetic ordering have been observed \cite{Hase,Masuda,Regnault,Fukuyama}. Field-induced and pressure-induced magnetic orderings are quantum phase transitions realized by the vanishing of the spin gap. The field-induced magnetic ordering can be interpreted as the Bose-Einstein condensation of triplet magnons \cite{Nikuni,Matsufield,Ruegg,Matsupaper}. The title spin gap system TlCuCl$_3$ undergoes all of these magnetic orderings \cite{Oosawamag,Oosawaheat,Tanakaela,Vyaselev,Sherman,OosawaMgsus,OosawaMgneu,TanakaLT,Oosawapressneu}. 

TlCuCl$_3$ has a monoclinic structure (space group $P2_1/c$) \cite{Takatsu}. The crystal structure of TlCuCl$_3$ consists of planar dimers of Cu$_2$Cl$_6$. The dimers form infinite double chains along the crystallographic $a$-axis. These chains are located at the corners and the center of the unit cell in the $b-c$ plane, and are separated by Tl$^+$ ions. The magnetic ground state of TlCuCl$_3$ is the spin singlet \cite{Takatsu} with the excitation gap $\Delta=7.7$ K \cite{Shiramura,Oosawamag}. The origin of the spin gap is the strong antiferromagnetic spin dimer in the chemical dimer Cu$_2$Cl$_6$, and the neighboring spin dimers couple via strong three-dimensional interdimer interactions along the double chain and in the $(1, 0, -2)$ plane, in which the hole orbitals of Cu$^{2+}$ spread \cite{Oosawainela,Cavadini}. 

\begin{figure}[t]
\begin{center}
\includegraphics[width=80mm]{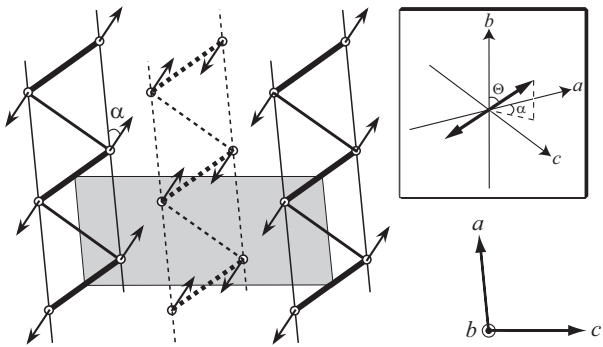}
\end{center}
	\caption{Projection of the spin structure observed in the pressure-induced magnetic ordered phase for $P=1.48$ GPa in TlCuCl$_3$ onto the $a-c$ plane. The double chains located at the corner and the center of the chemical unit cell in the $b-c$ plane are represented by solid and dashed lines, respectively. The shaded area is the chemical unit cell in the $a-c$ plane. The inset shows the inclination of the ordered spin toward the $b$-axis. The angle $\alpha$ denotes the angle between the $a$-axis and the spin component projected onto the $a-c$ plane. The angle $\Theta$ is the angle between the spin and the $b$-axis. }
	\label{spinstructure}
\end{figure}

TlCuCl$_3$ undergoes pressure-induced magnetic ordering above $P_{\rm c}\sim 0.2$ GPa \cite{TanakaLT}. Unpolarized neutron elastic scattering experiments under the hydrostatic pressure $P=1.48$ GPa were carried out in order to investigate the spin structure of the pressure-induced magnetic ordering in TlCuCl$_3$ \cite{Oosawapressneu}. Below the ordering temperature $T_{\rm N}=16.9$ K, the magnetic Bragg reflections were observed at reciprocal lattice points ${\mib Q}=(h, 0, l)$ with integer $h$ and odd $l$, which are equivalent to those points with the lowest magnetic excitation energy at ambient pressure. The spin structure of the pressure-induced ordered phase in TlCuCl$_3$ for $P=1.48$ GPa was determined as shown in Fig. \ref{spinstructure} with $\alpha=42.6^{\circ} \pm 1.4^{\circ}$ and $\Theta=90.0^{\circ} \pm 9.0^{\circ}$ for $T=12.2$ K and $\alpha=49.5^{\circ} \pm 2.4^{\circ}$ and $\Theta=58.0^{\circ} \pm 3.4^{\circ}$ for $T=4.0$ K. The spin structure for $T=12.2$ K is almost the same as those observed in the field-induced \cite{Tanakaela} and impurity-induced \cite{OosawaMgneu} magnetic ordered phases. This result implies that the ordered moments lying in the $a-c$ plane just below $T_{\rm N}$ incline toward the $b$-axis at a lower temperature. However, we could not determine from the previous unpolarized results whether the reorientation continuously occurs just below $T_{\rm N}$ or occurs because of the second phase transition at a lower temperature. Such inclination is directly observable using the polarized neutron scattering technique, because the spin components in and out of a scattering plane can be separated by measuring the spin-flip (SF) and non-spin-flip (NSF) scatterings from a single diffraction peak when the polarization direction is perpendicular to the scattering plane. It is also important to investigate whether the second phase transition, if it exists, is of the 1st order or 2nd order in order to discuss the origin of the second phase transition. Thus, in order to obtain clear answers to these questions, we carried out polarized neutron scattering experiments on TlCuCl$_3$ under hydrostatic pressure. 

The preparation of a single crystal TlCuCl$_3$ has been reported in ref. \citen{Oosawamag}. We used a 0.2 cm$^3$ sample set in a McWhan-type high-pressure cell \cite{McWhan} the same as that used in previous unpolarized experiments \cite{Oosawapressneu}. The applied hydrostatic pressure is $P=1.48$ GPa, the same as that in previous experiments \cite{Oosawapressneu}. Polarized neutron elastic scattering experiments using the uniaxial polarized neutron scattering technique \cite{Moon}, were performed using the JAERI-TAS1 installed at JRR-3M in Tokai. Heusler(111)-Heusler(111) monochromator-analyser configurations were used in the present experiments. The incident neutron energy $E_i$ was 14.7 meV. Because the size of the sample has to be small due to the small sample space in the high-pressure cell, collimators were set as open-80'-80'-80' in order to increase intensity. Sapphire and pyrolytic graphite filters were placed to suppress the background by high energy neutrons and higher order contaminations, respectively. The sample was mounted in the cryostat with its $a^*$- and $c^*$-axes in the scattering plane. The crystallographic parameters were determined as $a^*=1.6402$ $\rm{\AA}^{-1}$, $c^*=0.72843$ $\rm{\AA}^{-1}$ and $\cos\beta^*=0.0861$ at helium temperatures and $P=1.48$ GPa. In the present experiments, the guide field was applied perpendicular to the scattering $a-c$ plane, {\it i.e. }, vertical field (VF) configuration. In this configuration, the intensity of the SF scattering is proportional to the square of the spin component perpendicular to the scattering vector ${\mib Q}$ in the scattering $a-c$ plane $\langle S_{\perp}^{ac} \rangle^2$, while that of the NSF scattering is proportional to the square of the spin component perpendicular to both the scattering vector ${\mib Q}$ and the scattering $a-c$ plane, namely to the square of the $b$-axis component of spin $\langle S^{b} \rangle^2$ for a magnetic Bragg reflection. The polarization was measured on the nuclear Bragg reflection and was determined to be approximately 70\%. This value is considerably worse than the polarization of almost 90\% realized in the standard polarization analysis setup on JAERI-TAS1. This is due to the ferromagnetic components of the high-pressure cell. We checked that the polarization does not change in the experimental temperature range, {\it i.e.}, independent of the high-pressure cell temperature.

\begin{figure}[t]
\begin{center}
\includegraphics[width=80mm]{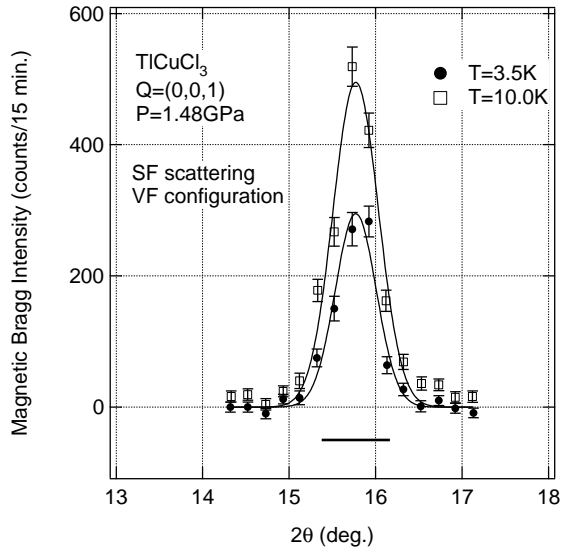}
\end{center}
	\caption{Magnetic Bragg intensities for the SF scattering at ${\mib Q} = (0, 0, 1)$ measured at $T=3.5$ K and 10.0 K for $P=1.48$ GPa in TlCuCl$_3$. The solid lines and horizontal bar denote the fittings by a Gaussian function and the calculated instrumental resolution width, respectively.}
	\label{Q001profile}
\end{figure}

Figure \ref{Q001profile} shows the magnetic Bragg intensities for the SF scattering at ${\mib Q} = (0, 0, 1)$ measured at $T=3.5$ K and 10.0 K for $P=1.48$ GPa in TlCuCl$_3$. The intensities are corrected for the background measured above $T_{\rm N}$. For $T=3.5$ K and 10.0 K, which are lower than $T_{\rm N}=16.9$ K, the Bragg intensity attributed to the magnetic ordering can be clearly seen, as observed in the previous unpolarized experiments \cite{Oosawapressneu}. It should be noted that the magnetic Bragg intensity at $T=10.0$ K is larger than that at $T=3.5$ K. This indicates that the additional change in spin structure occurs at low temperatures in the ordered phase. The Bragg reflection retains resolution limited width at these temperatures, as shown in Fig. \ref{Q001profile}.  

\begin{figure}[t]
\begin{center}
\includegraphics[width=80mm]{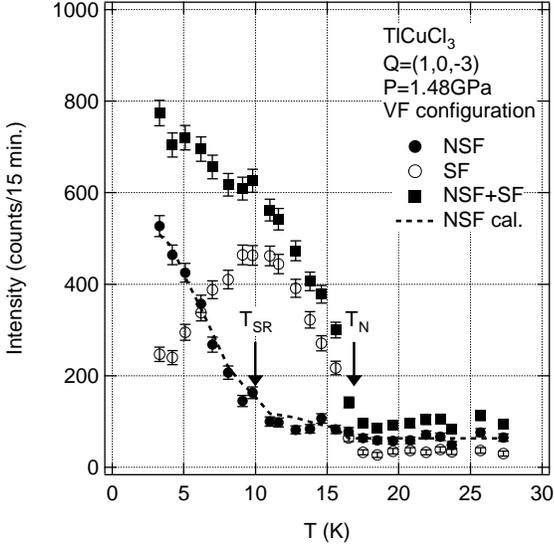}
\end{center}
	\caption{Temperature dependence of the Bragg peak intensities of SF and NSF scatterings, and of the total Bragg peak intensity of both scatterings for the ${\mib Q}=(1, 0, -3)$ reflection measured at $P=1.48$ GPa in TlCuCl$_3$. The dashed line denotes the temperature dependence of the calculated NSF scattering intensity (see in the text).}
	\label{Q10-3temdep}
\end{figure}

Figure \ref{Q10-3temdep} shows the temperature dependence of the Bragg peak intensities of SF and NSF scatterings, and of the total Bragg peak intensity of both scatterings for ${\mib Q}=(1, 0, -3)$ reflection measured at $P=1.48$ GPa. The increase of the magnetic Bragg intensity can be clearly seen below $T_{\rm N}=16.9$ K, as observed in the previous unpolarized experiments \cite{Oosawapressneu}. The peak intensity of the SF scattering increases, while that of the NSF scattering retains the background level just below $T_{\rm N}$ with decreasing temperature. The small intensity increase of the NSF scattering just below $T_{\rm N}$ can be accounted for by the incomplete polarization mentioned above. However, at $T_{\rm SR}=10$ K, the peak intensity of the SF scattering has a maximum, and then decreases. On the other hand, the peak intensity of the NSF scattering begins to increase below $T_{\rm SR}$. Because all the observed intensities at ${\mib Q}=(1, 0, -3)$ below $T_{\rm N}$ can be safely assigned to be of magnetic origin, the intensities of the SF and NSF scatterings in the present experiment are proportional to the square of the spin component perpendicular to the scattering vector ${\mib Q}$ in the scattering $a-c$ plane $\langle S_{\perp}^{ac} \rangle^2$ and that parallel to the $b$-axis $\langle S^{b} \rangle^2$, respectively. This result indicates that the ordered moments which lie in the $a-c$ plane just below $T_{\rm N}$ begin to incline toward the $b$-axis below $T_{\rm SR}$. Because the peak intensity of the NSF scattering shows a continuous and rapid increase at $T_{\rm SR}$, we can deduce that the inclination arises from a phase transition of the 2nd order. We also confirmed that the temperature dependence shows no hysteresis. As shown in Fig. \ref{Q10-3temdep}, the total intensity of the SF and NSF scatterings which can be observed in unpolarized experiments has no clear anomaly indicative of the phase transition at $T_{\rm SR}$, as observed in the previous unpolarized experiments \cite{Oosawapressneu}. We also confirmed that the temperature dependence of the total intensity coincides with that of the magnetic Bragg intensity at ${\mib Q}=(1, 0, -3)$ obtained in the previous unpolarized experiments (see Fig. 3 in ref. \citen{Oosawapressneu}), when multiplied by a scale factor. The phase transition at $T_{\rm SR}$ could be observed using the present polarized neutron scattering technique for the first time. In the previous magnetization measurements under the hydrostatic pressure for $P < 0.8$ GPa, the phase transition corresponding to $T_{\rm N}$ in the present measurement was detected for $P > P_{\rm c}\sim 0.2$ GPa, and no additional transition was observed down to $T=1.8$ K \cite{TanakaLT}. Thus, we infer that the second phase transition occurs for $P > 1$ GPa. 

\begin{figure}[t]
\begin{center}
\includegraphics[width=80mm]{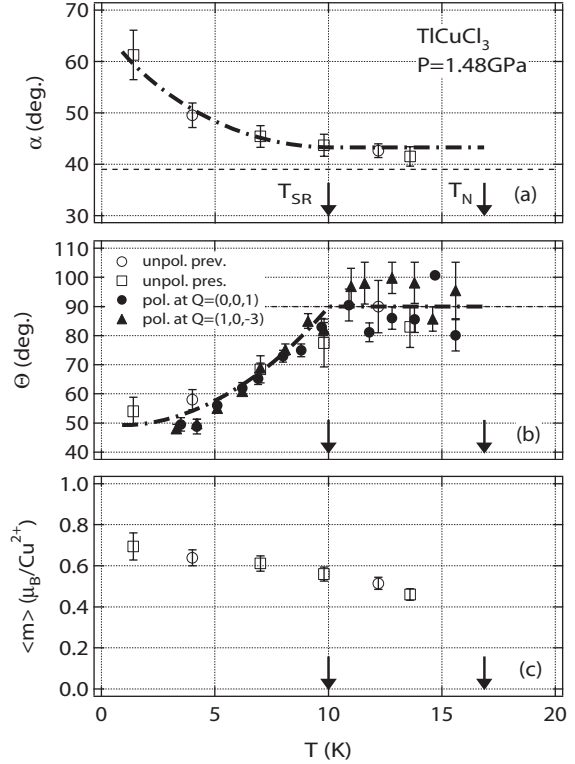}
\end{center}
	\caption{Temperature dependence of (a) $\alpha$, (b) $\Theta$ and (c) the magnitude of the ordered moment $\langle m \rangle$ at $P=1.48$ GPa. The closed circles and triangles denote the angles $\Theta$ determined from the ratio of SF and NSF magnetic Bragg peak intensities at ${\mib Q}=(0, 0, 1)$ and $(1, 0, -3)$ corrected for the guide field tilt angle, respectively. The open circles and squares denote the values estimated in the previous \cite{Oosawapressneu} and present unpolarized experiments, respectively. The dashed lines indicate the angles observed in the field-induced \cite{Tanakaela} and impurity-induced \cite{OosawaMgneu} magnetic ordered phases. The dotted-dashed lines are guides for the eyes.}
	\label{angletemdep}
\end{figure}

The angle $\Theta$ can be evaluated as $34.8^{\circ} \pm 1.5^{\circ}$ from the ratio of SF and NSF magnetic Bragg peak intensities for ${\mib Q}=(1, 0, -3)$ at $T=4.0$ K, as shown in Fig. \ref{Q10-3temdep}. The discrepancy between this value and $\Theta=58.0^{\circ} \pm 3.4^{\circ}$ estimated from the previous unpolarized experiments \cite{Oosawapressneu} indicates that the applied guide field is tilted slightly from the $b$-axis due to the ferromagnetic components of the high-pressure cell. Assuming that the tilting angles of the guide field $\psi$ and $\phi$ are 16$^{\circ}$ and 21$^{\circ}$, in which $\psi$ and $\phi$ are the angle between the guide field and the $b$-axis, and that between the $a$-axis and the guide field projected onto the $a$-$c$ plane, respectively, the temperature dependence of the angle $\Theta$ can be obtained from the ratio of SF and NSF Bragg peak intensities at ${\mib Q}=(0, 0, 1)$ and $(1, 0, -3)$ as shown in Fig. \ref{angletemdep} (b). We can clearly see that the angle $\Theta$ which is 90$^{\circ}$ just below $T_{\rm N}$ begins to decrease gradually below $T_{\rm SR}$. The observed temperature dependence of the NSF scattering at both ${\mib Q}$-values can be consistently explained under these assumptions, as shown in Fig. \ref{Q10-3temdep} in the case of ${\mib Q}=(1, 0, -3)$. It is noted that the angle $\alpha$ is fixed at 45$^{\circ}$ in this calculation for the simplification of the calculation because the calculated values are almost independent of the angle $\alpha$ estimated in this temperature range above $T=4$ K, namely from $\alpha=40^{\circ}$ to 50$^{\circ}$. It is also noted that the polarizations had to be assumed to be 68 \% and 90 \% for ${\mib Q}=(0, 0, 1)$ and $(1, 0, -3)$ scattering configurations, respectively. The scattering configuration dependence of the polarization can be attributed to the field inhomogeneities in the neutron path due to the ferromagnetic components of the high-pressure cell, as mentioned above. \par

We also carried out {\it un}polarized neutron scattering experiments at several temperatures under the same instrumental conditions as the previous experiments \cite{Oosawapressneu}. The results obtained are shown in Fig. \ref{angletemdep}. These values are estimated using the same analyses as the previous experiments \cite{Oosawapressneu} with the reliability factors of $R=\sum_{h,k,l}|I_{\rm cal}-I_{\rm obs}|/\sum_{h,k,l}I_{\rm obs} = 0.06 \sim 0.10$. As shown in Fig. \ref{angletemdep}, the temperature dependence of the angle $\Theta$ obtained in the unpolarized experiments is consistent with that obtained in the polarized experiments. The temperature dependence of the magnitude of the ordered moment $\langle m \rangle$ is also plotted in Fig. \ref{angletemdep}. This shows no anomaly at $T_{\rm SR}$ and increases monotonically with decreasing temperature. It is noted that the temperature dependence of $\langle m \rangle$ does not necessarily correspond to that of the total magnetic Bragg peak intensity, because the angles are dependent on temperature. 

When the $x$-direction is taken to be parallel to the ordered moments, the basis state of the dimer $\psi$ may be expressed as 
\begin{equation}
\psi=\left|0,0\right> \cos{\theta}+\frac{1}{\sqrt{2}}\left(\left|1,1\right> - \left|1,-1\right> \right)\sin{\theta}\ , 
\end{equation}
where angle $\theta$ was introduced to satisfy the normalization condition \cite{OosawaMillion}. From eq. (1), the magnitude of sublattice magnetization $\langle m \rangle$ is given by $\langle m \rangle=g\mu_{\rm B}\cos{\theta}\sin{\theta}$. With $\langle m \rangle\approx 0.70\ \mu_{\rm B}$ for $T\rightarrow 0$ and $g\approx 2.1$, we obtain $\sin^2{\theta}\approx 0.13$. This value denotes the contribution of the triplet component to the ground state at $P=1.48$ GPa.

Successive magnetic phase transitions have been observed in the frustrated spin systems which undergo three and more sublattice magnetic orderings, such as triangular antiferromagnets, in order to release the frustration partially \cite{Collins}. However, the successive phase transitions in the present system seem not to be explained by this case because the present system undergoes two-sublattice collinear magnetic ordering, as shown in Fig. \ref{spinstructure}, though there is a small frustration in the exchange network \cite{Oosawainela}. Also, the successive phase transitions accompanied by the inclination of ordered spin moment have been observed in mixed spin systems as Fe$_{1-x}$Co$_x$Cl$_2$ \cite{Wong,Mukamel}, Fe$_{1-x}$Co$_x$Cl$_2\cdot$2H$_2$O \cite{Katsumata} and CsCu$_{1-x}$Co$_x$Cl$_3$ \cite{Ono}. In these mixed systems, the inclinations of ordered spin moments have been discussed in terms of the competing random anisotropies or the off-diagonal exchange interaction. However, the successive phase transitions in the present system without disorder may not be explained by competing anisotropies of the quadratic form because in pure system they cannot produce the gradual change of spin direction, but produce sudden change. The competition between the quadratic anisotropy with the prefered axis in the $a-c$ plane and the anisotropy of the fourth order with the preferred axis along the $b$-axis can stabilize the intermediate spin direction. Since the magnitude of spin is $S=\frac{1}{2}$ in the present system, the intrinsic anisotropy of the fourth order seems not to be comparable with the quadratic anisotropy. The energy of the magnetoelastic coupling is effectively described by the fourth order coupling between sublattice magnetic moments, and the presence of the coupling was actually demonstrated by the NMR experiment \cite{Vyaselev}. Therefore, it is suggested that the magnetoelastic coupling is responsible for the inclination of orderd spin moments below $T_{\rm SR}$.

In conclusion, we have presented the polarized neutron elastic scattering results on the spin gap system TlCuCl$_3$ under the hydrostatic pressure $P=1.48$ GPa. Pressure-induced successive magnetic orderings were observed at $T_{\rm N}=16.9$ K and $T_{\rm SR}=10$ K. The second phase transition at $T_{\rm SR}$ was detected by the present polarized neutron experiment for the first time. The ordered moments lie in the $a-c$ plane below $T_{\rm N}$. However, at $T_{\rm SR}$ the ordered moments starts to incline toward the $b$-axis. The temperature variations of the direction and the magnitude of the ordered moments were evaluated as shown in Fig. \ref{angletemdep}. To our best knowledge, the present result is the first observation of pressure-induced successive magnetic phase transitions in quantum spin systems {\it without disorder} which undergo quantum phase transitions accompanied by {\it two-sublattice collinear} magnetic ordering.

We thank Y. Shimojo for his technical support. This work was supported by the Toray Science Foundation and a Grant-in-Aid for Scientific Research on Priority Areas (B) from the Ministry of Education, Culture, Sports, Science and Technology of Japan. \par

\end{document}